# Spin Solid versus Magnetic Charge Ordered State in Artificial Honeycomb Lattice of Connected Elements

*Artur Glavic, Brock Summers, Ashutosh Dahal, Joseph Kline, Walter Van Herck, Alexander Sukhov, Arthur Ernst, and Deepak K. Singh\**

The nature of magnetic correlation at low temperature in two-dimensional artificial magnetic honeycomb lattice is a strongly debated issue. While theoretical researches suggest that the system will develop a novel zero entropy spin solid state as $T \rightarrow 0$ K, a confirmation to this effect in artificial honeycomb lattice of connected elements is lacking. This study reports on the investigation of magnetic correlation in newly designed artificial permalloy honeycomb lattice of ultrasmall elements, with a typical length of ≈12 nm, using neutron scattering measurements and temperature-dependent micromagnetic simulations. Numerical modeling of the polarized neutron reflectometry data elucidates the temperature-dependent evolution of spin correlation in this system. As temperature reduces to ≈7 K, the system tends to develop novel spin solid state, manifested by the alternating distribution of magnetic vortex loops of opposite chiralities. Experimental results are complemented by temperature-dependent micromagnetic simulations that confirm the dominance of spin solid state over local magnetic charge ordered state in the artificial honeycomb lattice with connected elements. These results enable a direct investigation of novel spin solid correlation in the connected honeycomb geometry of 2D artificial structure.

## 1. Introduction

Two dimensional artificial honeycomb lattice provides a facile platform to explore many novel properties of magnetic materials in one system.[1,2] It includes the ice analog of magnetism, spin ice, spin liquid, and an unusual spin solid state, depicted by the distribution of magnetic vortex loops of opposite chiralities.[3,4] The complex variety of entropy controlled magnetic phases that are predicted to arise in an artificial honeycomb lattice as a function of reducing temperature cannot be realized in a "3D" bulk material of geometrically frustrated origin. According to recent theoretical reports, the honeycomb lattice behaves as a paramagnet at high temperature, corresponding to a gas of ±1 and ±3 magnetic charges.[5,6] As temperature is reduced, the system crosses over into a spin-ice type state, manifested by "two-in & one-out" (or vice-versa) configuration where two of the moments along the honeycomb element point to the vertex and one points away from the vertex (or vice-versa). At further reduction in temperature, a new ordering regime, characterized by the topological "charge order" of ±1 magnetic charges, develops (depicted in Figure 3). Thermal energy is expected to be comparable to the strength of the dipolar interaction (≈D) in the charge ordered regime. The transition from a local spin ice to the charge ordered state is chiral in nature, as a number of mobile closed loops of each chirality develop.[3,5,6] At much lower temperature, the system is predicted to evolve into a "spin-ordered" state of chiral vortex loops with zero entropy density, also called the "spin solid" state. It

Dr. A. Glavic
Laboratory for Neutron Scattering and Imaging
Paul Scherrer Institut
5232, Villigen PSI, Switzerland

B. Summers, A. Dahal, Prof. D. K. Singh
Department of Physics and Astronomy
University of Missouri
Columbia, MO 65211, USA
E-mail: singhdk@missouri.edu

Dr. J. Kline
National Institute of Standards and Technology
Gaithersburg, MD 20899, USA

Dr. W. Van Herck
Jülich Centre for Neutron Science (JCNS) at
Heinz Maier-Leibnitz Zentrum (MLZ)
Forschungszentrum Jülich GmbH
Lichtenbergstr. 1, 85748 Garching, Germany

Dr. A. Sukhov
Forschungszentrum Jülich GmbH
Helmholtz Institute Erlangen-Nürnberg for Renewable Energy (IEK-11)
90429, Nürnberg, Germany

Dr. A. Ernst
Institut für Theoretische Physik
Johannes Kepler Universität
A 4040, Linz, Austria

Dr. A. Ernst
Max-Planck-Institut für Mikrostrukturphysik
Weinberg 2, 06120 Halle, Germany

represents a novel phase of magnetic matter with zero entropy and magnetization.[7]

In this letter, we report the investigation of the spin solid state in artificial honeycomb lattice of connected ultrasmall elements. The typical dimension of a connecting element is 12 nm (length) × 5 nm (width) × 10 nm (thickness). Detailed measurements of polarized neutron reflectometry and glancing incidence small angle neutron scattering have revealed the development of additional scattering of magnetic origin from in-plane correlations when cooling down to $T = 7$ K. The diffuse band-type scattering is reasonably well described by the numerical modeling of the spin solid state configuration where magnetic moments along the connecting permalloy elements of the honeycomb lattice manifest an alternating order of the vortex loops of opposite chiralities. Development of the spin solid state is independently confirmed by the temperature-dependent micromagnetic simulations that show a temperature-dependent evolution of the underlying spin correlation in an artificial honeycomb lattice with similar size of connecting elements. The system tends to attend this novel order as $T \to 0$ K.

The current experimental efforts in accessing the spin solid state in artificial honeycomb lattice is mainly based on the employment of the electron-beam lithography method for sample fabrication, which generally leads to small sample sizes with large elemental parameters. Such large element size honeycomb lattice typically manifests an inter-elemental interaction energies of ≈$10^4$ K.[1,8–10] More recently, a new design of artificial honeycomb lattice, consisting of very thin (few angstroms) and well-separated permalloy elements of large size (≈500 nm in length and 20–50 nm in width) that reduces the interelemental energy significantly, was proposed to access the spin solid state in the disconnected geometry.[11,12] In the disconnected honeycomb lattice, the underlying physics is dictated by the Block transition mechanism. On the other hand, the physical connection between the lattice elements in connected honeycomb lattice facilitates the propagation of domain wall from one vertex to another, which is necessary for inducing the spin flip and, hence, the continuous progression of a new magnetic phase as temperature is reduced.[8,13,14] Here, we take a different approach and create large throughput new artificial honeycomb lattice of connected ultrasmall elements. The ultrasmall size of the connecting elements automatically reduces the interelemental magnetostatic energy between 12 and 15 K. Therefore, it is quite suitable for the investigation of temperature-dependent evolution of magnetic phases using reciprocal space probes, such as neutron scattering method.

## 2. Results and Discussion

The fabrication of artificial honeycomb lattice samples involved the synthesis of hexagonal diblock copolymer templates and near parallel deposition of permalloy material on top of the honeycomb structured silicon substrates in an ultrahigh vacuum chamber. Similar diblock copolymer templates are extensively used to fabricate nanostructured materials.[15] Under suitable physical conditions, a diblock copolymer tends to self-assemble where one component tends to develop long-range periodic structures.[16,17] Additionally, the flexibility in tuning the structural properties and lattice parameters, by simply varying the composition and molecular weight of the diblock copolymer, allow to create a plethora of nanomaterials.[18,19] Some of the notable examples include the fabrication of nanodot, nanoring, and nanoparticle assemblies.[15,17,19,20] More recently, researchers have used diblock templates and glancing angle deposition to create directional hierarchical structures of metal nanoparticles.[17] An atomic force microscopy image of a typical honeycomb sample is shown in **Figure 1**a (see the Experimental Section for detail). A small angle X-ray scattering measurement in the grazing incidence angle configuration (GISAXS) confirms the high structural quality of the sample. GISAXS measurements can provide information about the structural properties of a system.[21–23] The GISAXS measurements were performed using a Ga K source with a wavelength of 1.34 Å at an incident angle of 0.15°. A 1 mm thick stainless steel foil was used to attenuate the reflected beam. As shown in Figure 1b, GISAXS measurements show a primary spacing of 31 nm, which is consistent with the atomic force microscopy image within the calibration error. The second and third peaks visible in the scattering pattern occur at multiples of $\sqrt{3}$ and 2 of the primary peak, corresponding to a 2D hexagonal lattice. The higher order peaks seem to be overshadowed by the background in the data, which is most likely arising due to the possible inhomogeneity in the sample. The dimension of the constituting element of the honeycomb lattice is not perfect, rather varies a little bit from the average size of 12 nm in length and 5 nm in width. However, the small variation in the element size is not expected to affect the underlying physics much, as the interelemental energy will only change marginally (less than 2 K for a variation of up to 2 nm). From the modeling of the GISAXS data, the large domain size of long-range structural order in the honeycomb lattice was confirmed (paracrystal correlation length of 250 nm).

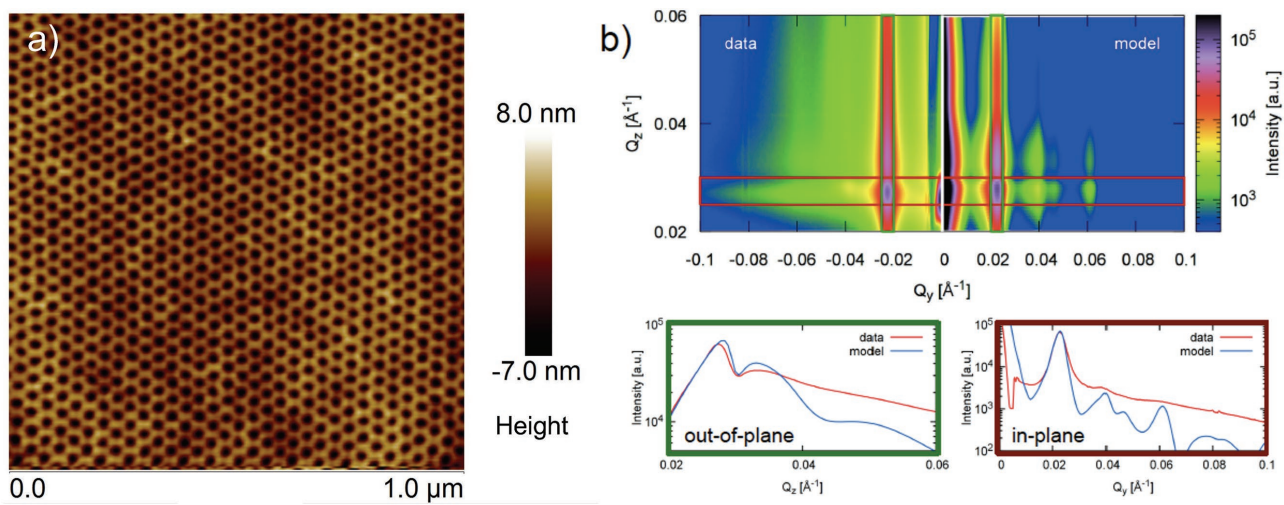

**Figure 1.** Structural characterization of artificial honeycomb lattice. a) Full size atomic force microscopy image of typical artificial honeycomb lattice, derived from diblock porous template combined with reactive ion etching (see the text for detail). The bond length, width, and lattice separation are ≈12, 5, and 31 nm, respectively. b) Grazing incident X-ray scattering recorded with an incidence angle of 0.15° using Ga $K_\alpha$. 2D plots, as shown below, are horizontal and vertical integrations of the areas marked as red and green boxes in the image. Numerical simulations, using the same structural parameters as for the neutron models (discussed below), are shown in the same graph for comparison and describe the main features and their positions accurately.

To directly investigate the correlation between magnetic moments along the honeycomb elements, we have performed polarized neutron experiments, namely reflectometry, off-specular and grazing incidence neutron scattering (GISANS) measurements.[21,24–26] Together, they allow us to explore the underlying magnetic correlations[27] at a length scale of ≈5 nm to 10 µm in the honeycomb lattice. In **Figure 2**, we plot the off-specular intensity measured using spin-up (+) and spin-down (−) neutrons at $T = 300$ K and 7 K. Here, the y-axis represents the out-of-plane scattering vector $\left(Q_z = \frac{2\pi}{\lambda}(\sin\alpha_i + \sin\alpha_f)\right)$ while, the difference between the z-components of the incident and the outgoing wave vectors $\left(p_i - p_f = \frac{2\pi}{\lambda}(\sin\alpha_i - \sin\alpha_f)\right)$ is drawn along the x-axis. Thus, vertical and horizontal directions correspond to the out-of-plane and in-plane correlations, respectively (for detail information, see Lauter et al.[28]). The specular reflectivity lies along the $x = 0$ line. Clearly, the off-specular scattering plots show remarkable differences between high and low temperature measurements. At $T = 300$ K, the specular intensity is more than two orders of magnitude stronger compared to the off-specular data, which is the expected behavior for most systems. There is also some scattering in the off-specular regions caused by the paramagnetic nature of the moment and the honeycomb structure itself. The difference between the spin-up and the spin-down components in the off-specular reflections is similar to that observed in the specular data (see Figure S1, Supporting Information). It indicates a paramagnetic or weak ferromagnetic character (MSLD = $5.9 \times 10^{-6}$ Å$^{-2}$ ↔ $M = 1.92 \times 10^5$ A m$^{-1}$) of the honeycomb lattice at room temperature. Upon cooling the sample to $T = 7$ K, the off-specular signal increases significantly (notice the logarithmic color scale). No specular beam can be distinguished from the off-specular background, anymore. As the nuclear structure factor will not change noticeably upon cooling, this can only be explained by a significant change in the magnetic characteristics of the system.

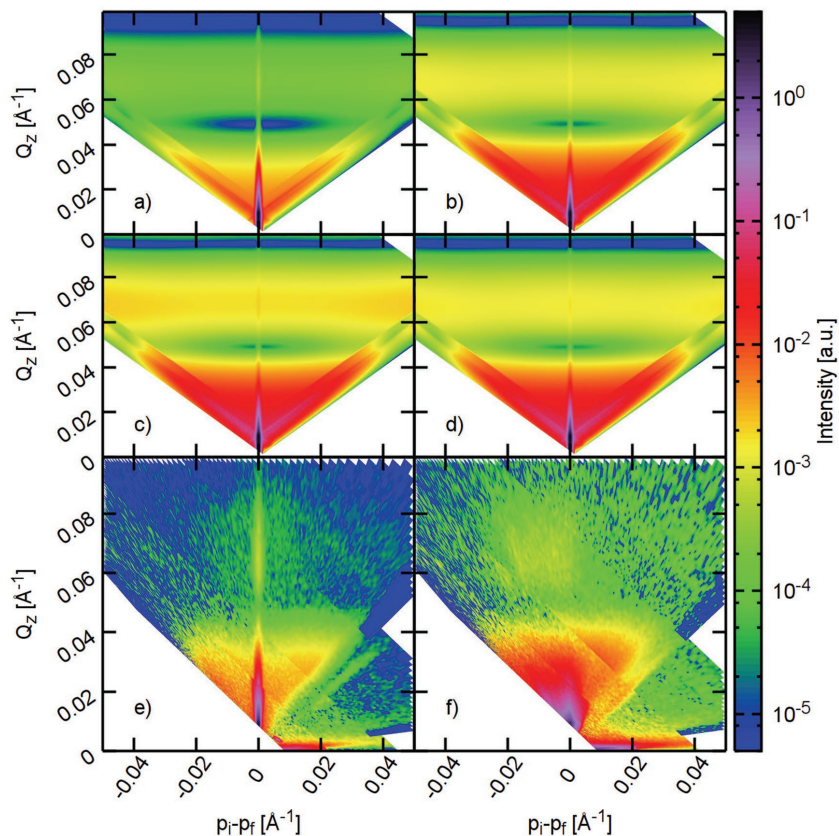

**Figure 2.** Polarized neutron scattering measurement on artificial honeycomb lattice. a–d) Off-specular neutron reflectometry simulated in DWBA for the various models shown as sum of all neutron spin states. Starting with a) a paramagnetic spin-gas state, the correlations of spins (see Figure 3) increase up to d) the spin solid state. The corresponding polarized neutron data recorded at e) $T$ 300 K and f) 7 K are shown as sum of spin-up and spin-down components. In all graphs, the x- and y-axes indicate the difference ($p_i - p_f$) and sum ($Q_z$) of the out-of-plane components of the incident and outgoing neutron wave vectors that correspond to lateral and depth correlations, respectively. The specular reflection at room temperature, indicating the paramagnetic state (e), is replaced by a broad diffuse scattering, indicating correlation due to ≈100 nm structure size (f). Unlike the paramagnetic case, the simulated spin solid state exhibits strong scattering along the horizontal axis of similar magnitude as the specular reflectivity. The simulated profiles well describe the experimental results. The steps in the data represent change in the incidence angle, typical of a time of flight (TOF) measurement.

The broad feature along the horizontal axis in the neutron reflectometry data at $T = 7$ K indicates the development of in-plane magnetic correlations in the artificial honeycomb lattice (Figure 2). In order to further understand the underlying magnetic structure, experimental data is compared to the numerically simulated patterns for theoretically predicted magnetic phases, namely paramagnetic state as well as spin ice (ice-1), charge ordered configuration (ice-2) and spin solid states, all depicted in **Figure 3**. The different magnetic states were simulated in our Distorted Wave Born Approximation (DWBA) model (see the Experimental Section for more information).[29,30] As shown in the lower panel of Figure 2, the off-specular scattering increases with the amount of spin–spin correlation in the system. Although the difference between ice-2 (magnetic charge ordered state) and spin solid is small, a reasonable agreement with the experimental data is obtained for the spin solid state. The spin solid state is modeled by arranging the vortex loops of opposite chiralities in an alternating order, as described above. The simulated patterns indeed exhibit broad bands of diffuse scattering along the x-axis, as observed in the experimental data (also see the Supporting Information for detail about the modeling). We note that the interelemental energy in the newly designed artificial honeycomb lattice is ≈12 K. This is the characteristic energy necessary for a magnetic charge ordered state to develop, followed by the spin solid state as $T \to 0$ K. Therefore, the observed increase in intensity is consistent with the expected behavior in our sample.

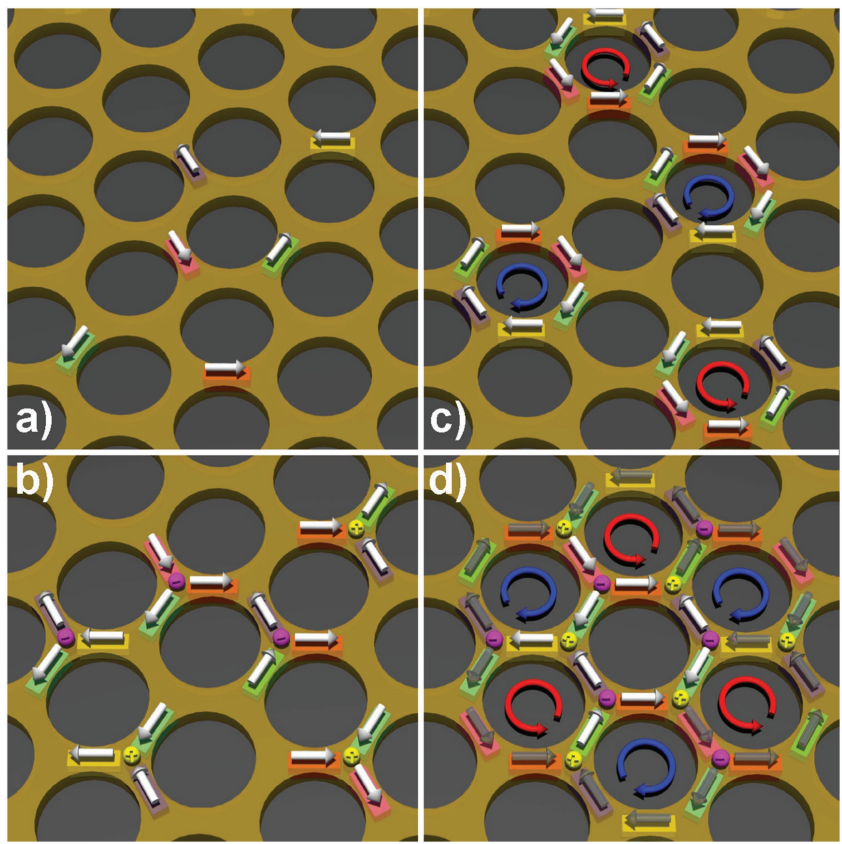

**Figure 3.** Magnetic correlation models used for the DWBA simulations. While in all models the number of spins on the lattice is equal to the number of honeycomb edges, the correlations are limited to a) single spins for the paramagnetic, b) spin triangles with 2-in & 1-out or 1-in & 2-out arrangements for ice-1, c) spin vortices of either left or right chiralities for ice-2, and d) long range ordered arrangement of chiral vortex loops for the spin solid state. Pink and yellow balls at the vertex represent ±1 unit of magnetic charge.

To further understand the magnetic behavior at low temperature, we have also performed GISANS experiments with limited intensity available, as the beamline is not optimized for this method. The signal-to-background ratio is not large enough to observe distinguishable features in the reciprocal space map (see Figure S2, Supporting Information). We have performed integrations along the $Q_y$ axis using a band from $Q_z = 0.025$ Å$^{-1}$–0.045 Å$^{-1}$ for the models depicted in Figure 3 and the measured data, which are shown in **Figure 4**. The 300 K data have a clear peak around $Q_y = 0.02$ Å$^{-1}$, which corresponds to the nuclear structure as well as the total scattering in the gas and ice-1 states. Upon cooling down to 7 K, additional intensity develops around $Q_y \approx 0.012$ Å$^{-1}$ that is only expected for the ice-2 and/or the spin solid states (also see Figures S3–S5, Supporting Information for detail simulations). For the sample to manifest an Ice-2 state, a finite intensity at $Q_y \approx 0.025$ Å$^{-1}$ should have been observed. Rather, it is missing in the experimental data. The observed intensity profile, albeit limited, is reasonably consistent with the predicted q-dependence of intensity in the spin solid state or, a phase mixture of spin solid state (majority phase) with ice states (minority phase).

Next, we have performed temperature-dependent micromagnetic simulations on an artificial honeycomb lattice of similar element size and thickness to independently verify the development of a spin solid state at low temperature. The honeycomb lattice was simulated using 0.2 nm × 0.2 nm mess size on the $\mu$MAG platform,[31,32] with magnetic field applied in-plane to the lattice. The simulation utilizes the Landau–Lifshitz–Gilbert equation of magnetization relaxation in a damped medium. It is given by[33–35]

$$\frac{d\mathbf{m}}{dt} = -\gamma \mathbf{m} \times \mathbf{h}_{\text{eff}} + \alpha \mathbf{m} \times \frac{d\mathbf{m}}{dt}, \quad (1)$$

where $\gamma$ is the gyromagnetic ratio and $\alpha$ is damping constant. In the above equation, thermal fluctuation is introduced by Langevin dynamics.[36] The Langevin dynamics utilizes the concept of white noise of Gaussian form, $\Gamma(t)$ with a mean at zero, arising due to the thermal fluctuation. Accordingly, each site experiences a white noise as temperature increases. At each time step, the instantaneous thermal field on each site is given by, $\mathbf{h}_{\text{therm}} = \Gamma(t)\sqrt{\frac{2k_B T \alpha}{\gamma \mu_s \Delta t}}$, where $\mu_s$ is magnetic moment of simulation element $i$. The thermal field adds to the original effective field via $\mathbf{h}(T > 0)_{\text{eff}} = \mathbf{h}(T = 0)_{\text{eff}} + \mathbf{h}_{\text{therm}}$, where $h_{\text{eff}}(T = 0) = -\delta \mathbf{H}/\delta \mathbf{m}$. The Hamiltonian, **H**, of the system consists of four terms: exchange energy, uniaxial anisotropic energy, magnetostatic energy and the Zeeman energy. For the simulation, we have used exchange stiffness $A = 1.0 \times 10^{-11}$ J m$^{-1}$, uniaxial anisotropy strength $K_1 = -1.0 \times 10^3$ J m$^{-3}$, and damping constant $\alpha = 0.2$. The simulated hysteresis curves at various temperatures are shown in **Figure 5**. Qualitative differences between magnetic hysteresis curves at different temperatures are clearly observed in this figure. The plot of $M/M_s$ versus $h$, where $M_s$ is saturation magnetization, depicts a more subtle transition in applied field as temperature reduces. At low temperature (simulation at $T = 0$ K), magnetic hysteresis not only manifests a sharp transition near zero field but also develops a small plateau near $M = 0$. The simulated magnetization profile in this state is characterized by the vortex configuration, which is the key element of the spin solid state. Unlike the development of vortex state near zero field at low temperature, the magnetization profile at $T = 100$ K depicts short-ranged ordered spin ice state (see the color map). The micromagnetic simulations further confirm the development of spin solid state.

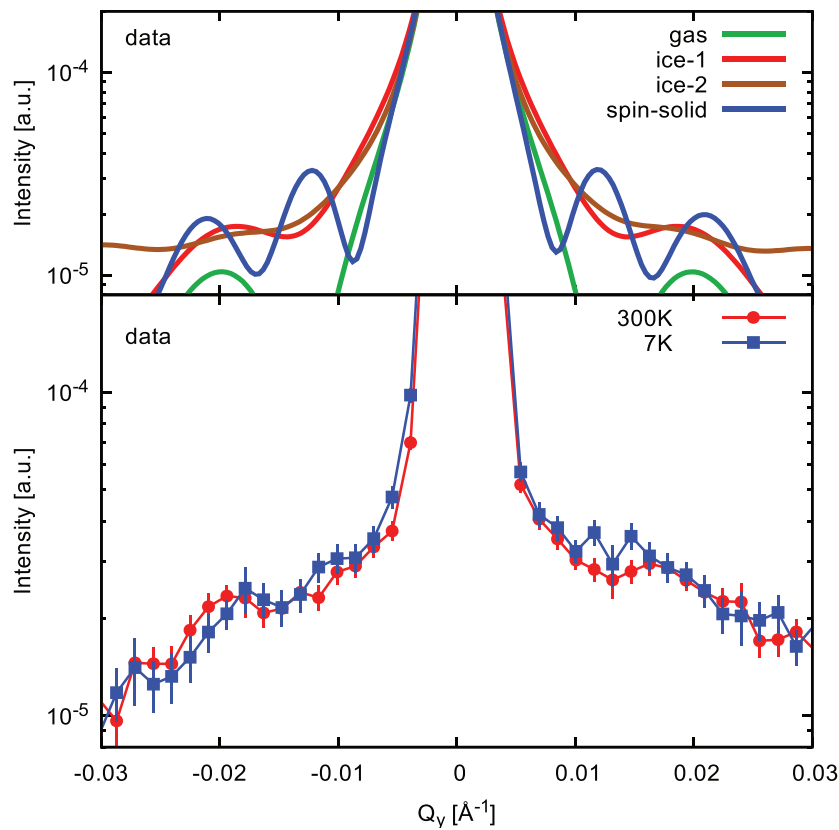

**Figure 4.** Cuts through the simulated (top) and measured (bottom) GISANS maps on a range from $Q_z = 0.025$ Å$^{-1}$ to 0.045 Å$^{-1}$. There is an increase in intensity around $Q_y \approx 0.012$ Å$^{-1}$ that corresponds well to the simulated increase in the spin-solid state. While there is an increase in the ice-2 state for this region as well, the additional intensity at 0.03 Å$^{-1}$ is not observed in the experiment. The initial 300 K state could well be a mixture of spin-gas and ice-1 state.

## 3. Conclusion

In summary, we have presented experimental investigation of magnetic correlations at low temperature in newly designed artificial honeycomb lattice. The experimental results were independently verified by the temperature-dependent micromagnetic simulations. Among the various magnetic phases that are expected to arise as a function of reducing temperature in artificial honeycomb lattice, magnetic charge ordered state and spin solid state hold greater significances.[37] Both states are somewhat unique to this 2D structure that involve chiral vortex loops. While the charge ordered state is expected to develop below magnetostatic dipolar interaction temperature, given by $D \approx k_B T$, the spin solid state arises as $T \to 0$ K. Based on our experimental and theoretical researches, we infer that the magnetic moments along the honeycomb element in the newly fabricated honeycomb lattice tend to develop the spin solid state, compared to magnetic charge ordered state (ice-2 phase), as temperature reduces below the interelemental energy, $T \approx 12$ K. Our results also suggest the highly competing nature of novel magnetic phases in artificial honeycomb lattice of connected elements. Future experimental researches, such as the estimation of entropy per element, are desirable to further understand the development of the spin solid state[6,7] A real time imaging technique, such as Lorentz microscopy, can provide direct evidence of spin solid state in this system. Future efforts in this regard are specially desirable.

## 4. Experimental Section

*Sample Fabrication and Characterizations*: Fabrication of artificial honeycomb lattice involves the synthesis of porous hexagonal diblock template on top of a silicon substrate, calibrated reactive ion etching using $CF_4$ gas to transfer the hexagonal pattern to the underlying silicon substrate, and the deposition of magnetic material (permalloy) on top of the uniformly rotating substrate in near-parallel configuration ($\approx 1°$) to achieve the 2D character of the system. The sample fabrication process utilized diblock copolymer polystyrene(PS)-*b*-poly-4-vinyl pyridine (P4VP) of molecular weight 23K Dalton with the volume fraction of 70% PS and 30% P4VP. The self-assembly of diblock copolymer was driven by microphase separation arising from the immiscibility of the polymer blocks. A microphase separated diblock copolymer film can take various forms from spherical to cylindrical to lamellar, depending upon the volume fraction of each block. At this volume fraction, the diblock copolymer tends to self-assemble, under right condition, in a hexagonal cylindrical structure of P4VP in the matrix PS.[38] A 0.5% PS-*b*-P4VP copolymer solution in toluene was spin casted onto cleaned silicon wafers at 2500 rpm for 30 s and placed in vacuum for 12 h to dry. The samples were solvent annealed at 25 °C for 12 h in a mixture of THF/toluene (80:20 v/v) environment. The process results in the self-assembly of P4VP cylinders in a hexagonal pattern within a PS matrix. The average diameter of a P4VP cylinder was ≈ 12 nm and the center-to-center distance between two cylinders was ≈ 30 nm, also consistent with that reported by Park et al.[38] Submerging the samples in ethanol for 20 min releases the P4VP cylinders yielding a porous hexagonal template. The diblock template was used as a mask to transfer the topographical pattern to the underlying silicon substrate. The top surface of the reactively etched silicon substrate resembles a honeycomb lattice pattern. This property is exploited to create metallic honeycomb lattice by depositing permalloy, $Ni_{0.81}Fe_{0.19}$, in near parallel configuration in an electron-beam evaporation. For this purpose, a new sample holder was designed and setup inside the e-beam chamber. The substrate was rotated uniformly about its axis during the deposition to create uniformity. This allowed evaporated permalloy to coat the top surface of the honeycomb only, producing the desired magnetic honeycomb lattice with a typical element size of 12 nm (length) × 5 nm (width) × 8 nm (thickness). Atomic force microscopy image of a typical honeycomb lattice is shown in Figure 1a (also see Figure S6 in the Supporting Information where it is shown that the average roughness in the thickness of a honeycomb element is less than 0.5 nm). The center-to-center spacing between neighboring honeycombs is ≈ 30 nm. Thus, each honeycomb is about 30 nm wide. Further details about the fabrication procedure can be found somewhere else. GISAXS, at an incident angle of 0.15°, confirmed the good structural quality of the sample.

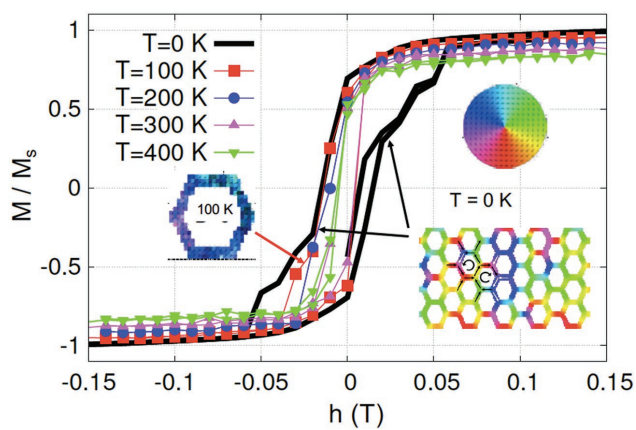

**Figure 5.** Temperature-dependent micromagnetic simulation. Micromagnetic simulations at $T = 0$, 100, 200, and 300 K show qualitative differences in magnetic hysteresis curves. A small plateau near zero field and magnetization at $T = 0$ K is identified with a magnetic configuration of chiral vortex loops, tending to form a spin solid state in connected honeycomb lattice. The simulated profile near zero field at $T = 100$ K exhibits spin-ice configuration. The chiral vortex loop disappears as temperature increases.

*Neutron Scattering Measurements*: Neutron scattering measurements were performed on a 25 mm × 25 mm surface area sample at the magnetism reflectometer, beam line 4A of the Spallation Neutron Source, at Oak Ridge National Laboratory. The instrument utilizes the time of flight technique in a horizontal scattering geometry with a bandwidth of ≈ 2.8 Å (wavelength varying between 2.2 and 5.0 Å). The beam was collimated using a set of slits before the sample and measured with a 2D position sensitive $^3$He detector with 1.5 mm resolution at 2.5 m from the sample. The sample was mounted on the copper cold finger of a close cycle refrigerator with a base temperature of $T = 7$ K. Beam polarization and polarization analysis were performed using reflective supermirror devices, achieving better than 98% polarization efficiency over the full wavelength band. For reflectivity and off-specular scattering, the full vertical divergence was used for maximum intensity and a 1% $\Delta\theta/\theta \approx \Delta Q_z/Q_z$ relative resolution in horizontal direction. These measurements were carried out with five *q*-ranges in a total of 380 min. For GISANS experiments, both directions were collimated to reach a symmetric resolution of 0.13° in $Q_y$ and $Q_z$ directions. Counting times for these measurements were 8 h per instrument setting (temperature, incident angle).

*Scattering Modeling*: The DWBA was used as implemented in BornAgain[29] to model the GISAXS, GISANS, and off-specular scattering intensity using the same model with different instrument and material parameters. The specular neutron reflectivity data at $T = 300$ K (see Figure S1, Supporting Information), fitted using the GenX software,[30] were used as a basis for the model generation. For more details about the used model see the Supporting Information and corresponding theoretical background in refs. [23,30,39–42].

## Supporting Information

Supporting Information is available from the Wiley Online Library or from the author.

## Acknowledgements

The research at MU was supported by the U.S. Department of Energy, Office of Basic Energy Sciences under Grant No. DE-SC0014461. A portion of this research used resources at the Spallation Neutron Source, a DOE Office of Science User Facility operated by the Oak Ridge National Laboratory.

## Conflict of Interest

The authors declare no conflict of interest.

## Keywords

artificial magnetic honeycomb lattices, geometrical frustration, neutron reflectometry measurements

Received: November 11, 2017
Published online: